# Ignition and Burn in a Small Magnetized Fuel Target

**Ronald C. Kirkpatrick,**

Los Alamos National Laboratory, Los Alamos, NM, USA

E-mail:   rck@lanl.gov

**Abstract**
LASNEX calculations of a small magnetized target show high gain at a velocity significantly lower than needed for unmagnetized targets.   Its cryogenic fuel layer appears to be raised to an equilibrium ignition temperature of about 2 keV by the radiation from the burning magnetized fuel.

## 1. Introduction

Ignition is the crucial step toward a sufficiently high gain to enable design of a power producing system based on inertial confinement fusion (ICF).   This has proven to be a daunting task, and despite a long-term, committed design effort, ignition of the NIF targets has yet to be experimentally demonstrated at the US DOE National Ignition Facility (NIF) [1] at Livermore, CA.

Magnetized target fusion (MTF) may offer an alternate approach that relaxes some of the most stringent constraints on driver capabilities and target fabrication.   The one dimensional (1-D) calculations presented here suggest that this approach may provide demonstration of fusion ignition in a NIF-like setting, as well as the gain and other characteristics needed for a practical fusion reactor.

This paper focuses on the physical processes in a small magnetized fusion target that enables it to ignite a cold fuel layer and thereby provide high gain.

## 2. Background

Over three decades ago a series of small scale experiments at Sandia National Laboratory first explored the benefit of magnetizing a small fusion target [2].   The Sandia e-beam target consisted of a spherical microballoon with a very thin collector plate mounted on the cathode side was connected to the anode of the electron beam machine by a stalk.   Charge that collected on the plate during a non-relativistic prepulse discharged through the microballoon, creating a diffuse z-pinch that both preheated and magnetized the fuel in side.   The powerful relativistic pulse that followed imploded the microballoon.   Post-shot analysis [3] of the series of 25 shots indicated that a magnetic field was essential for explaining the neutron production from these "Phi-targets."   The target presented below have a similar basic geometry, but differ significantly. Not long after the series of   Phi-target experiments, Sweeney and Farnsworth performed calculations that suggested a suitable design might provide high gain [3a].   During the intervening years there have been several numerical surveys and studies (e.g., [4-6]) and a few experimental endeavors (e.g., [7-10a]) exploring various aspects of what is now known as magnetized target fusion.

## 2. Basic physics

The basic idea of MTF is to use a suitable magnetic field embedded in the fusion plasma to a) reduce the electron thermal conduction and b) enhance the fusion energy deposition. This requires 1) operation at lower density than used for typical laser-driven ICF target designs, as well as 2) provision of a means of magnetizing and preheating the low density fuel. The main advantage of magnetization is reduction of the compression rate (therefore, implosion velocity) necessary to achieve fusion conditions. The reduced velocity opens up the possibility of using efficient, energetic pulsed power facilities to deliver the drive necessary to achieve ignition [5]. The calculations presented here suggest that it may also be possible to demonstrate the basic physics of MTF on a laser facility such as the NIF.

In order to reduce the electron thermal conduction, it is necessary have a sufficiently large product of the collision time and cyclotron frequency for the electrons. A closed field configuration is preferable for low implosion velocities (e.g., $< 5$ cm/$\mu$s). In addition, in order to enhance self-heating by charged fusion products it is necessary to have a Larmor radius for the charged fusion products that is smaller than the size of the target at the time of maximum compression. Whereas ICF requires the areal density ($\rho R$) to be greater than ~ 0.3 gm/cm$^2$ [6], in a magnetized target the field-times-radius parameter (BR) augments $\rho R$, so that $\rho R$ can be much smaller than 0.3 gm/cm$^2$. The threshold for augmentation is BR ~ 0.3 Tm (MGcm) during fusion burn. While electron thermal conduction is the major energy transport mechanism for typical low density ICF targets [11], upon reducing conduction by imposing a magnetic field, radiation from a low density magnetized fuel becomes the dominant transport mechanism. Thus far, a few analytic studies and "zero-D" numerical surveys have provided a bit of insight, but more is needed.

## 3. Lasnex calculations

In order to understand the how the various physical processes interact, the Lasnex ICF simulation code [12] was used for a parameter study of a particular target configuration. This target configuration is based on the results of a previous study [13]. The modified target used here consists of three spherical shells (aluminum, gold, and frozen deuterium-tritium) enclosing a low density gas with an annular magnetic field due to an axial current. The thin opaque layer of gold traps the radiation generated by the burning magnetized fuel. In all the calculations thus far, the three solid shells were started with a uniform inward velocity of 10 cm/$\mu$s, which is considerably lower than the 40 cm/$\mu$s required for current NIF targets [1]. The DT gas is initially at rest, so the gas experiences an initial shock, which raises its temperature. The 1-D calculation was done as an equatorial wedge rather than the usual cone-on-axis model typical of 1-D calculations for laser-driven ICF targets. This facilitates an initial azimuthal (B-theta) magnetic field of about 10 T. Substantial gain occurs over a fairly wide range of magnetized fuel initial densities and temperatures, as shown in Figures 2 and 3.

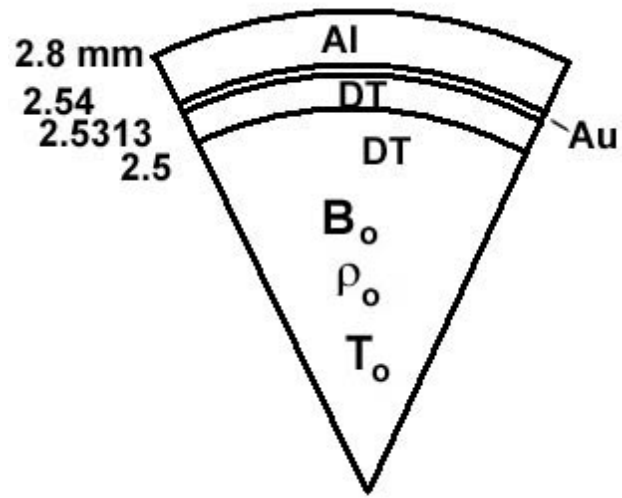

Figure 1. Configuration of the target.

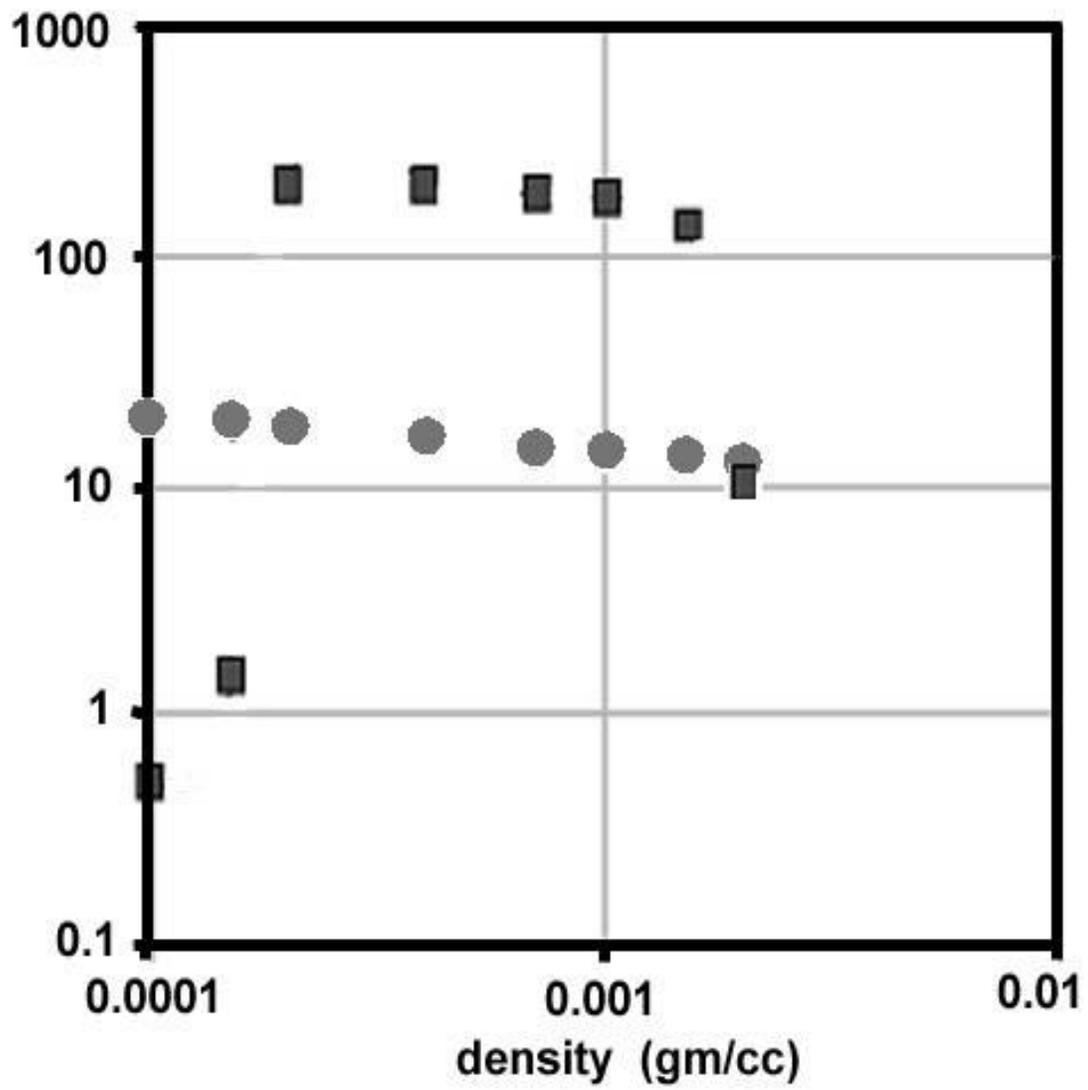

Figure 2. Gain (squares) and convergence (circles) vs initial magnetized fuel density.

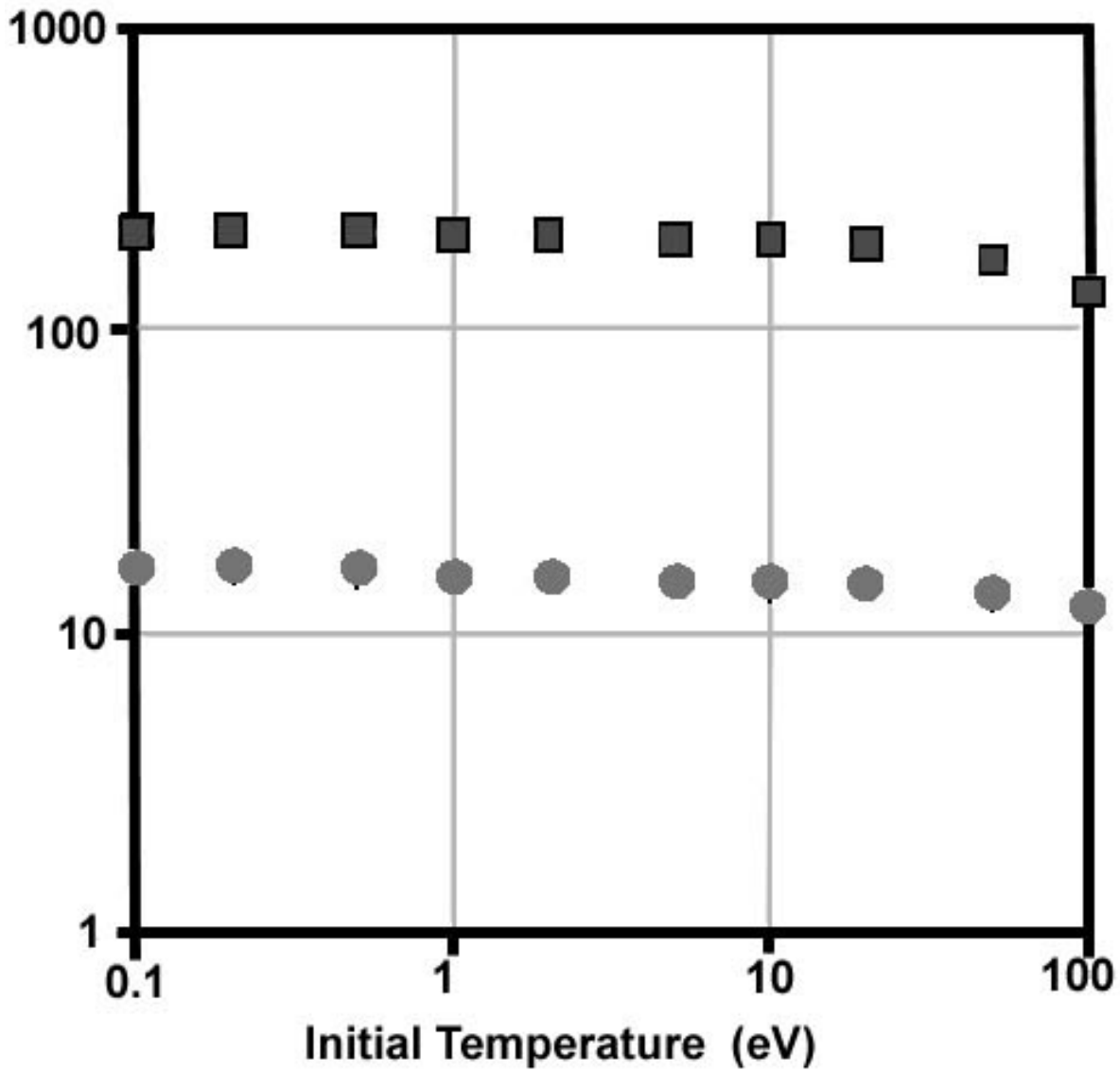

Figure 3. Gain (squares) and convergence (circles) vs initial magnetized fuel temperature.

It should be noted that the gains for these 1-D calculations may be optimistic by perhaps a factor of about two, because the fusion burn near the axis in a 2-D calculation is not expected to proceed to the same degree as that in the equatorial region. It is remarkable that the gain and convergence do not seem to suffer as the initial temperature in the magnetized fuel decreases. One might conjecture that this is due to the initial shock propagating into the DT is able to sufficiently ionize it and allow trapping of the initial magnetic field. Answers to this and other questions must await further analysis.

**4. Surprising Physics**

The most remarkable aspect of these calculations is the mechanism whereby the burning magnetized fuel ignites the cold fuel layer. Jones and Mead showed that supressed thermal conductivity makes it difficult to propagate burn from a hot spot into a magnetized fuel [14]. However, the calculations reported here show that a) the radiation is trapped inside the thin gold layer and b) the cold fuel undergoes classic equilibrium ignition [11], which is quite different from the propagating burn responsible for lighting the cold fuel in an ICF target. While the magnetized fuel ignites in a volume ignition mode, if the burn is sufficiently intense, it produces sufficient radiation to then raise the cold fuel to the equilibrium ignition temperature. It is necessary to raise the radiation temperature above 2 keV, so that as the cold fuel comes into equilibrium with the radiation it achieves ignition. The effect of the magnetic field on the high density cold fuel appears to be minimal, but the synergism between the two fuel regions is crucial.

## 5. Conclusions

It may be possible to achieve high gain in a NIF-sized magnetized fuel target. However, the calculations presented here are far from a target design. First, they are idealized, having set initial conditions with no indication of how to achieve them. Second, the initial kinetic energy exceeds the absorbed energy available in a NIF hohlraum. However, the reduced velocity may allow a NIF reconfiguration to deliver the required energy. Third, the stability of the target needs to be assessed, and the effect of impurities in the DT fuel have not been addressed. If the target configuration presented here were to be fielded at NIF, it may be necessary to scale it to a smaller size, which may not be possible, or at least would likely impact its performance. However, it may be possible to demonstrate ignition of just the magnetized fuel with a lower implosion velocity, hence lower absorbed energy.

Statistics from the Omega (Rochester) and Nova (Livermore) lasers presented as plots of YOC (yield divided by calculated) versus convergence ratio (e.g., Fig. 4 of Varnum, et al., [15]) show that targets with high convergence ratios perform poorly. The target presented here provides high gain with a fairly low convergence ratio. Also, low convergence and low initial aspect ratio greatly relax the required drive symmetry and target fabrication tolerances.

There are potentially various approaches for preheating and magnetizing the target. Whichever is adopted must be capable of acting sufficiently quickly to leave the cold fuel layer intact, yet not significantly spoil the symmetry of the target. These approaches need further study, to be followed up with detailed 2-D calculations.

## 6. Summary

The calculations presented here suggest that a small target with magnetized fuel can ignite a cold fuel layer and provide sufficient gain to make such a target the basis for a fusion reactor study, which should spur interest in magnetized targets. Additional research is needed to more thoroughly explore the vast parameter space and improve the computational physics applied to this problem.

**Acknowledgements**

The author greatly appreciates the help, advice and support of Ian Tregillis, Bob Watt, Scott Hsu, Erick Lindman, and Charles Knapp. Also, this work would not have been possible without Los Alamos National Laboratory's generous Guest Agreement program.**References**

[1]    Lindl, J D, Atherton1, L J, Amednt, P A, Batha, S, Bell, P, Berger, R L, Betti, R, Bleuel, D L, Boehly, T R, Bradley, D K, et al., 2011 *Nucl. Fusion* **51** 094024.

[2]    Widner, M M, 1977 *Bulletin of the American Physical Society* **22** 1139**;** Olsen, J.N., Widner, M M, Chang, J, and Baker, L, 1979   J. Appl. Phys. **50** 3224.

[3]    Lindemuth, I R, and Widner, M M, 1981 *Physics of Fluids* **24** 753.

[3a] Sweeney M.A., and Farnsworth, A., 1981 *Nucl. Fusion* **21**, 41.

[4]    Lindemuth, I R, and   Kirkpatrick, R C, 1983 *Nucl. Fusion* **23**, 263.

[5]    Kirkpatrick, R C, Lindemuth, I R, Reinovsky, R E, and Ward, M S, 1992 in Megagauss Magnetic Field Generation and Pulsed Power Applications, Cowan, M, and Spielman, R B, editors, Nova Science Pub., Inc., Commack, NY.

[6]    Kirkpatrick, R C, Lindemuth, I R, and Ward, M S, 1995 *Fusion Technology* **27**, 201.

[7]    Gotchev, O V, Jang, N W, Knauer, J P, Barbero, M D, Betti, R, Li, C K, and Petrasso, R D, 2008 *Journal of Fusion Energy* **27** 25.

[8]    Intrator, T, Sears, J, Turchi, P J, Waganaar, W J, Weber, T, Wurden, G A, Degnan, J H, Domonkos, M, Grabowski, C, Ruden, E L, et al., 2010 *American Physical Society*, 52nd Annual Meeting of the APS Division of Plasma Physics, November 8-12, abstracts #UP9.098-105 and #UP9.099-105. http://adsabs.harvard.edu/abs/2010APS..DPPUP9098I

[9]    Fiksel, G., Chang, P.Y., Hohenberger, M., Knauer, J.P., Betti, R., Marshall, F.J., Meyerhofer, D D, S'eguin, F H, and Petrasso, R D, 2011 *Bulletin of the American Physical Society* **56** UO8.8.

[10]   Hsu, S C, Awe, T J, Brockington, S, Case, A, Cassibry, J T, Kagan, G, Messer, S J, Stanic M, Tang, X, Welch, D R, and Witherspoon, F D, 2012 *IEEE Trans. Plasma Sci.* 40, 1287.

[10a] Slutz, S, and Vesey, R A,   2012 *Physical Review Letters* **108**, 025003 (http://prl.aps.org).

[11]   Kirkpatrick, R C, and Wheeler, J A, 1982 *Nucl. Fusion* **21**, 389.